%% file: main.tex
\documentclass[floats,floatfix,showpacs,amssymb,prl,twocolumn,superscriptaddress,nofootinbib,nolongbibliography,reprint,aps]{revtex4-2}
\usepackage{textcomp}
\usepackage{amssymb,amsmath,verbatim,mathtools,needspace,enumitem,etoolbox,graphicx,microtype,afterpage,bigints,gensymb,tabularx,xspace}
\usepackage{aasmacros}
\usepackage{graphicx}%
\usepackage{dcolumn}%
\usepackage{bm}%
\usepackage{lipsum}
\usepackage[dvipsnames, table]{xcolor}
\usepackage{physics}
\usepackage{soul}

\definecolor{linkcolor}{rgb}{0.0,0.3,0.5}
\usepackage[colorlinks = true,
            linkcolor = linkcolor,
            urlcolor  = linkcolor,
            citecolor = linkcolor,
            anchorcolor = linkcolor]{hyperref}  
\usepackage{orcidlink}
\usepackage{array} 

\definecolor{linkcolor}{rgb}{0.0,0.3,0.5}
\usepackage[colorlinks = true,
            linkcolor = linkcolor,
            urlcolor  = linkcolor,
            citecolor = linkcolor,
            anchorcolor = linkcolor]{hyperref}

\usepackage[draft,defaultcolor=teal,commentmarkup=footnote, commandnameprefix=always]{changes}

\usepackage[capitalize]{cleveref}
\Crefname{equation}{Eq.}{Eqs.}
\Crefname{figure}{Fig.}{Figs.}
\Crefname{tabular}{Tab.}{Tabs.}
\Crefname{section}{Sec.}{Secs.}
\Crefname{subsection}{Sec.}{Secs.}
\Crefname{Appendix}{App.}{Apps.}

\usepackage{array}

\definechangesauthor{RB}

\newcommand{\pyring}{\textsc{pyRing}\xspace}
\newcommand{\pseob}{\textsc{pSEOB}\xspace}
\newcommand{\granita}{\textsc{granita}\xspace}

\usepackage{orcidlink}
\begin{document}

\title{
Functional inference on deviations from General Relativity}

\newcommand{\bham}{\affiliation{Institute for Gravitational Wave Astronomy \& School of Physics and Astronomy, University of Birmingham, Birmingham, B15 2TT, UK}}
\newcommand{\milan}{\affiliation{Dipartimento di Fisica ``G. Occhialini'', Universit\'a degli Studi di Milano-Bicocca}}
\newcommand{\infn}{\affiliation{INFN, Sezione di Milano-Bicocca, Piazza della Scienza 3, 20126 Milano, Italy}}

\author{Costantino Pacilio~\orcidlink{0000-0002-8140-4992}}
\email{costantino.pacilio@unimib.it}
\milan
\infn
\author{Riccardo Buscicchio~\orcidlink{0000-0002-7387-6754}}
\milan
\infn
\bham

\date{\today}

\begin{abstract}
Extensions of general relativity often predict modifications to gravitational waveform morphology that depend functionally on source parameters, such as the masses and spins of coalescing black holes. However, current analyses of strong-field gravity lack robust, data-driven methods to infer such functional dependencies. In this work, we introduce \granita, a non-perturbative, theory-agnostic framework to characterize parameter-dependent deviations from general relativity using Gaussian process regression. Leveraging the flexibility of this method, we analyze both simulated data and real events from the LIGO-Virgo-KAGRA public catalog. We demonstrate the ability of our approach to detect and quantify waveform deviations across the parameter space. Furthermore, we show that the method can identify stochastic (non-deterministic) deviations, potentially arising from environmental effects or subdominant unmodeled physics. As gravitational-wave tests of strong gravity advance in precision, our framework provides a principled approach to constrain modified gravity and to mitigate contamination from astrophysical or instrumental systematics.
\end{abstract}

\maketitle

\noindent{\bf \em Introduction~--~}
Testing General Relativity (GR) in the strong field regime is a major scientific objective of gravitational-wave (GW) astrophysics, as a probe of extreme spacetimes and environments in the vicinity of black-hole (BH) horizons \cite{Berti:2018cxi,Berti:2018vdi}. 
GW detectors are sensitive to waveform morphology, which would be affected in multiple ways by gravitational physics beyond GR. 
Waveforms modifications can be captured by effective deviation parameters $\delta y$. When estimating individual event parameters, deviation parameters are inferred together with the GR ones or reconstructed in post-processing: the former corresponds to, e.g., parametrized post-Newtonian tests \cite{Arun:2006yw,Yunes:2009ke} or parametrized ringdown tests \cite{Gossan:2011ha,Meidam:2014jpa,Brito:2018rfr}, while the latter corresponds to, e.g., consistency tests \cite{Ghosh:2016qgn,Ghosh:2017gfp,Bhagwat:2021kfa}. All these tests are often referred to as ``agnostic'' because they do not assume \textit{a priori} a functional dependence of the deviations from the GR parameters describing each source. By contrast, theory-informed tests consider well-posed non-GR theories with known functional dependencies in the non-GR effects -- see \cite{Yunes:2016jcc} for a dictionary between agnostic and theory-informed tests.

Typically,  theory-informed tests are better posed and yield larger significance than agnostic ones \cite{Chua:2020oxn}. However, the scarcity of GW waveforms from well-defined theories beyond GR \cite{Okounkova:2022grv,Julie:2024fwy,Pierini:2022eim,Silva:2022srr,Maenaut:2024oci} poses a significant challenge to the application of theory-informed tests. For this reason, agnostic deviations are widely used as necessary conveniences to increase our sensitivity to unknown theories. Current approaches to testing GR with deviation parameters consider combining individual events hierarchically \cite{Zimmerman:2019wzo}, to characterize the distribution of $\delta y$ across multiple observations. 
For example, in~\cite{Isi:2019asy} the observed distribution $p(\delta y)$ is approximately parametrized as $\mathcal{N}(\mu,\sigma)$: the test infers $\mu$ and $\sigma$ hierarchically and checks whether they are consistent with zero deviations (but see \cite{Chua:2020oxn,Johnson-McDaniel:2021yge,Pacilio:2023uef} for interpretational caveats). Hierarchical inference has been applied to combine public GW observations by the LIGO-Virgo-KAGRA Collaboration \cite{LIGOScientific:2020tif,LIGOScientific:2021sio}. %

Robust methods to reconstruct the functional dependence between the deviation parameters and the GR parameters currently lack in literature.
As GW astronomy progresses towards precision science, the ability to not only \textit{detect} deviations from GR, but also \textit{characterize} their behavior across the parameter space becomes crucial to formulate theories consistent with observations \cite{Volkel:2022aca}.

Extending the scope of current tests in this direction is hampered by the lack of a general functional form to describe deviations from GR across the parameter space. Specific tests might admit a principled functional expression, albeit approximate or partial. For example, \cite{Payne:2024yhk} enhances hierarchical tests by explicitly parametrizing the dependence on the mass scale of the system, inspired by nonminimally coupled scalar-field theories, and with minimal assumptions about the type of tests that are performed. Restricted to BH spectroscopy \cite{Berti:2005ys,Berti:2025hly}, Ref.~\cite{Maselli:2019mjd} introduced \textsc{ParSpec}, a beyond-GR ringdown expansion perturbative in the remnant black-hole spin: while in principle the test can be made arbitrarily accurate by increasing the order of the expansion, in practice the expansion converges slowly and only the first few coefficients can be included without corrupting the test precision \cite{Carullo:2021dui,Maselli:2023khq}.

In this work, we take a major step forward in the general problem of characterizing functional deviations from GR, agnostically.
To this aim, we introduce \granita, a non-perturbative approach based on Gaussian processes \cite{williams2006gaussian}. We take advantage of their flexibility in modeling a wide variety of functional forms through kernel-based covariance functions, after observing data at specific locations called \textit{nodes} \cite{2020arXiv200910862W}. Our approach promotes the nodes to free parameters that can be estimated via hierarchical inference, much alike Ref.~\cite{Pozzoli:2023lgz} in a different context. 
We show that accurate reconstructions can be achieved even with a small number of nodes. Crucially, we also show that in regions of the parameter space with substantial data support, reconstruction by \granita is akin to one obtained by assuming a parametrized version of the true underlying model.

Additionally, departures from a deterministic functional form might arise due to environmental effects of astrophysical origin -- for example, accretion disks around BHs and dark-matter spikes \cite{Cole:2022yzw} --, to the presence of primary BH hair \cite{Herdeiro:2015waa}, or to  subdominant unmodelled physics.
In our work, we include an auxiliary variance parameter $\sigma$ to capture stochastic deviations from the explored functional forms. 
When testing GR, it is fundamental to disentangle systematic and environmental effects \cite{Barausse:2014tra,Gupta:2024gun, 2025PhRvD.111h4037R,2024PhRvL.132y1401C} from genuine GR deviations. 
Our strategy serves exactly this purpose: as we shall see,
in our formalism
the auxiliary variance parameter $\sigma$ can be interpreted as a degree of determinism, hence the presence of non-deterministic contributions can be identified and inferred upon.

\noindent{\bf \em Illustration of the method~--~}
In order to illustrate the method, we consider a toy model consisting of a deviation parameter $\delta y$ -- such that $\delta y=0$ when GR is true -- depending on a single GR parameter $\theta_{\rm GR}$. We draw $N=100$ values of $\theta_{\rm GR}$ from a normal distribution $\mathcal{N}(0.5,0.15)$, representative of an ensemble of observed GW events. We generate the corresponding values of $\delta y$ from a functional relation $\delta y = f_{\rm true}(\theta_{\rm GR})$, where
\begin{equation}
    \label{eq:y:1}
    f_{\rm true}(\theta_{\rm GR}) \!=\! a\left(\theta_{\rm GR}-0.5\right)\big[1+b\sin\big(2\pi(\theta_{\rm GR}-0.5)\big)\big]~\!\!,
\end{equation}
and we set $a=0.1$ and $b=0.5$. 
We choose purposefully this functional relation because it is irreducible to a finite polynomial expansion, so it better illustrates our non-perturbative functional reconstruction technique.

To reconstruct the functional dependence, we consider $N_{\rm nodes}=5$ equally spaced nodes in the interval $[0,1]$. While we keep their locations $\boldsymbol{X}_{\rm nodes}=\{X_1,\dots,X_{N_{\rm nodes}}\}$ fixed, we promote their values $\boldsymbol{Y}_{\rm nodes}=\{Y_1,\dots,Y_{N_{\rm nodes}}\}$ to free parameters of the model. For each query point $\theta_{\rm GR}$, the expected prediction for $\delta y$ is identified with the expected value of a Gaussian process, conditioned on the ordered couples $({X}_i,{Y}_i)$,
\begin{equation}
    \label{eq:hgpr:1}
    \mu_{\rm pred}(\theta_{\rm GR}|\boldsymbol{X}, \boldsymbol{Y})=\sum_{i,j=1}^{N_{\rm nodes}}k(\theta_{\rm GR},{X}_i)k^{-1}({X}_i,{X}_j){Y}_j~.
\end{equation}
The function $k(\cdot,\cdot)$ is the kernel that defines the Gaussian process. We adopt a squared-exponential kernel,
\begin{equation}
    \label{eq:kernel:1}
k({X}_1,{X}_2)=\exp\left(-\frac{({X}_1,{X}_2)^2}{2l^2}\right)
\end{equation}
where $l$ is a model hyperparameter called the kernel \textit{correlation length}. We fix it to $l=0.5$: we find that this choice suffices to recover accurate results for the cases considered in this work; however, in more general cases, different kernels and associated hyperparameters, e.g.~$l$, can be optimized through model selection -- for example, by comparing the Bayesian evidence of the model for different choices of $l$.

We allow the model to predict $\delta y$ spread over $\mu_{\rm pred}$ by a standard deviation $\sigma$, which is also treated as a free parameter. 
Therefore, $\delta y$ as a function of $\theta_{\rm GR}$ is drawn from a normal distribution,
\begin{equation}
    \label{eq:y:2}
{\delta y \mid \theta_{\rm GR},\boldsymbol{\alpha} \sim p(\delta y|\theta_{\rm GR},\boldsymbol{\alpha})=\mathcal{N}(\mu_{\rm pred}(\theta_{\rm GR}),\sigma)}\,.
\end{equation}

In the above equation, $\boldsymbol{\alpha}$ collectively represents the free parameters of the model $\boldsymbol{\alpha}=\{Y_1,\dots,Y_{N_{\rm nodes}},\sigma\}$.
We provide priors on $\boldsymbol{\alpha}$ in the
supplemental material.

Given observations ${D}=\{d_1,\dots,d_N\}$, the corresponding posterior for $\boldsymbol{\alpha}$ can be obtained from the hierarchical likelihood
\begin{widetext}
\begin{equation}
    \label{eq:likelihood:1}
    p(D|\boldsymbol{\alpha},\boldsymbol{\beta})=\prod_{i=1}^N \int d{\theta}_{\rm GR}\int d\delta y~p(d_i|{\theta}_{\rm GR},\delta y)p(\delta y|{\theta}_{\rm GR},\boldsymbol{\alpha})p({\theta}_{\rm GR}|\boldsymbol{\beta})
    \propto\prod_{i=1}^N\mathbb{E}_i\left[\frac{p(\delta y|{\theta}_{\rm GR},\boldsymbol{\alpha})p({\theta}_{\rm GR}|\boldsymbol{\beta})}{\pi_i({\theta_{\rm GR}},\delta y)}\right]
\end{equation}

\end{widetext}
where the expectation values $\mathbb{E}_i[\dots]$ are taken over the individual posteriors $p({\theta}_{\rm GR},\delta y|d_i)$, and $\pi_i({\theta_{\rm GR}},\delta y)$ denotes the prior used for the posterior recovery of the $i$-th observation.

In \eqref{eq:likelihood:1}, we have introduced the additional population parameters $\boldsymbol{\beta}$ to parametrize the observed distribution of $\theta_{\rm GR}$. In particular, in this illustrative example we assume $p(\theta_{\rm GR}|\boldsymbol{\beta})=\mathcal{N}(\mu_X,\sigma_X)$, with $\boldsymbol{\beta}=\{\mu_X,\sigma_X\}$ being the hyper-parameters with true values $\boldsymbol{\beta}_{\rm true}=\{0.5,0.15\}$.

We note here that a rigorous analysis should not only infer $\boldsymbol{\alpha}$ and $\boldsymbol{\beta}$ jointly \cite{Payne:2023kwj}, but also include selection effects \cite{2019MNRAS.486.1086M,Essick:2023upv}. Since selection effects are not expected to have a major impact on the cases considered in this work, we ignore them for simplicity. The interested reader is referred to the conclusions for a dedicated discussion. 

 \begin{figure}[t]
     \centering
          \includegraphics[width=\linewidth]{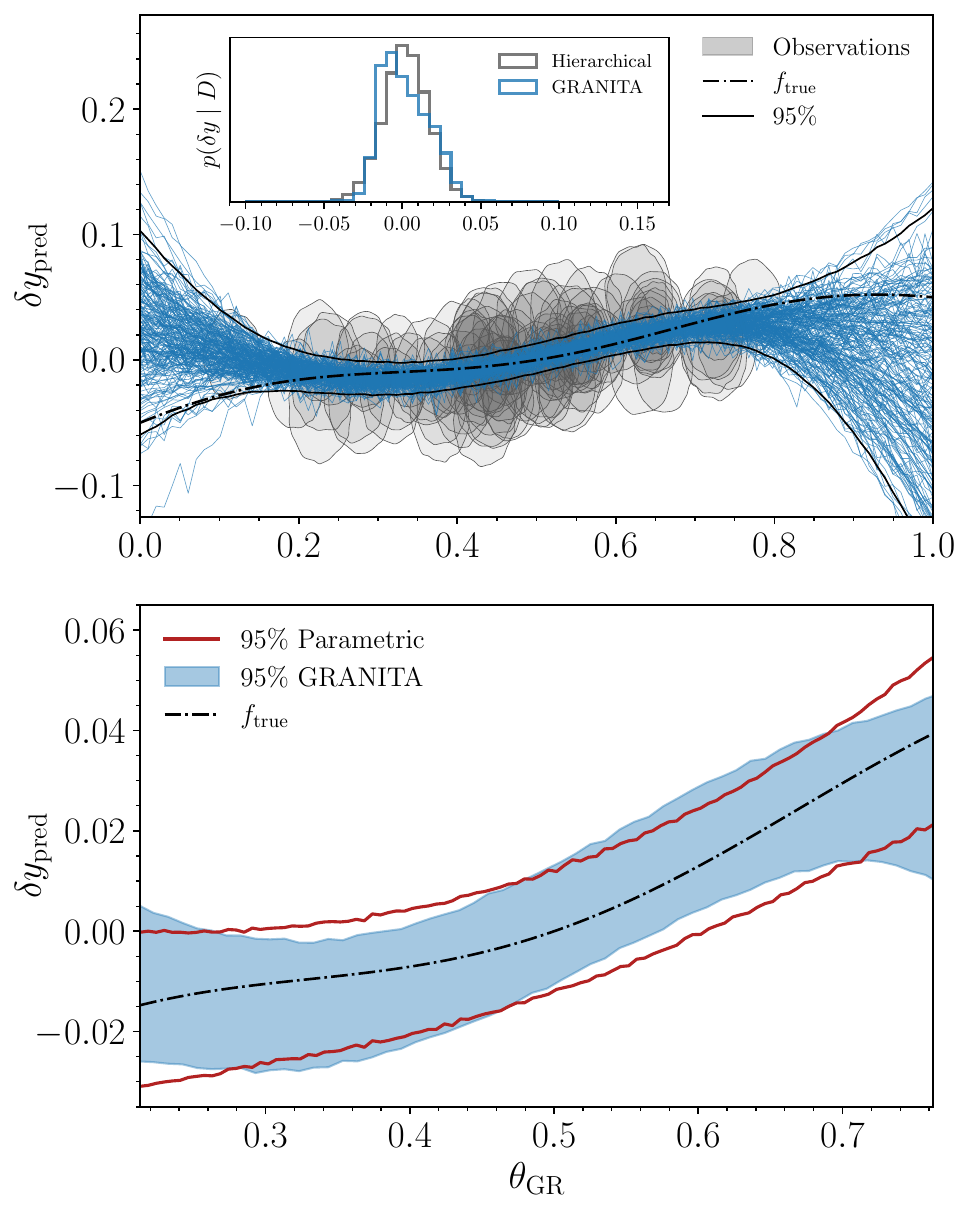}
     \caption{\textit{Top}: Reconstructed functional dependence of $\delta y$ on $\theta_{\rm GR}$, in the case that the true relation is given by Eq.~\eqref{eq:y:1}, with each blue line corresponding to an individual functional draw from the Gaussian Process \eqref{eq:y:2}. The gray contours show the 90\% credible regions of the mock observations considered in the simulated experiment. The inset show the posterior model distribution for $\delta y$, as predicted by \granita and by the hierarchical method. \textit{Bottom}: Comparison between the functional regions allowed by \granita and by the parametrized model \eqref{eq:y:3}, which contains the true model as a special case.}
     \label{fig:res:1}
 \end{figure}

We generate synthetic data assuming that the likelihood $\mathcal{L}(d_i|\theta_{\rm GR},\delta y)$ for each individual observation is a multivariate Gaussian with diagonal covariance and equal standard deviations $\sigma_{\rm obs}=1/\rho$ for $\theta_{\rm GR}$ and $\delta y$, where $\rho$ is the signal-to-noise ratio (SNR), assumed to follow a density distribution $p(\rho)\propto\rho^{-4}$ in the interval $[50,100]$. We add a Gaussian scatter to mimic the stochastic biases due to the presence of noise. Finally, we assume that the priors $\pi_i$ are flat, equal for all observations, and wider than the likelihood support, so that we can neglect them in~\cref{eq:likelihood:1}.

Figure \ref{fig:res:1} (\textit{Top}) shows the reconstructed dependence of $\delta y$ on $\theta_{\rm GR}$. We see that the individual functional draws encompass the true function $f_{\rm true}$ from Eq.~\eqref{eq:y:1} across the whole domain of $\theta_{\rm GR}$. In regions with observational support, the 95~\% credible band provides a much tighter constraint along the $y$ direction, compared to the values spanned by the contours of the individual events; at the same time, the functional uncertainties become wide in regions with few or no data.

The inset shows the posterior distribution for $\delta y$: it has very similar support and is broadly compatible with the corresponding prediction from the hierarchical test, the differences being due to the fact that the two methods employ different approximation schemes to model the deviation parameters.

Figure \ref{fig:res:1} (\textit{Bottom}) shows a comparison between the functional support allowed by \granita, and the one allowed by a parametrized model that contains the true functional relation \eqref{eq:y:1}: in the latter case, we model the expected value of $\delta y$ by
\begin{align}
    \label{eq:y:3}
    \!\!\mu_{\rm par}(\theta_{\rm GR})\! &=\!\!A\left(\theta_{\rm GR}\!-0.5\right)\!\left[1\!+\!B\sin\!\left(2\pi C(\theta_{\rm GR}\!-0.5)\right)\right]~\!\!\!,
\end{align}
and we sample the model parameters $\boldsymbol{\alpha}'=\{A,B,C,\sigma\}$ from the likelihood \eqref{eq:likelihood:1}, setting priors as specified in
the supplemental material.
We restrict the comparison to the region of $\theta_{\rm GR}$ that is most supported by the data -- specifically, to the $95\%$ posterior support of $\theta_{\rm GR}$ -- where the credibility of the prediction is more meaningful. Remarkably, we see that \granita achieves prediction bounds that are comparable to those of the parametrized model, which contains the true model as a special case. This is a highly desirable feature, since a parametrized model containing the true underlying functional dependence is almost always inaccessible. We take this as a further confirmation that the \granita method is efficient in exploring unspecified functional forms while retaining small posterior uncertainties.

In a second illustrative experiment, we extend the model by introducing a stochastic contribution to the generative model for $\delta y$, in the form of
\begin{equation}
    \label{eq:y:4}     
    \delta y_{\rm true} = f_{\rm true}(\theta_{\rm GR}) +\varepsilon Q,\quad\text{with}~\varepsilon\sim\mathcal{N}(0,0.025)\,.
\end{equation}
Here, $Q$ is an effective parameter describing additional system properties, e.g. its interaction with the environment or nuisance effects. For concreteness, we take $Q=\theta_{\rm GR}^2$, although our findings have general validity. While our model \eqref{eq:y:2} only allows for a scatter of order $\sigma$ about $\mu_{\rm pred}$, the stochastic contribution in \eqref{eq:y:4} cannot be reduced to a draw from normal distribution. Therefore, this setting offers a non-trivial benchmark for our method.

\begin{figure}[t]
    \centering
    \includegraphics[width=\linewidth]{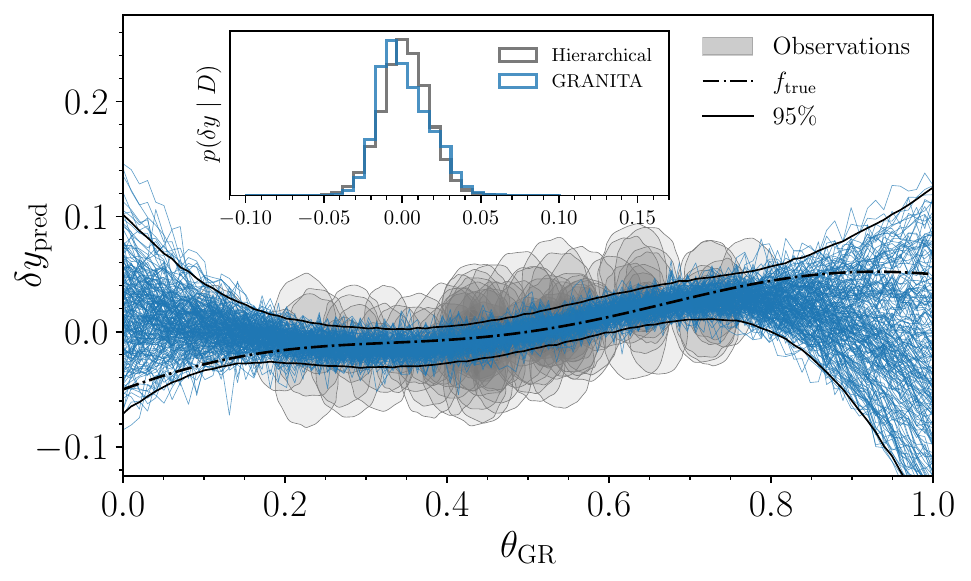}
    \caption{Same as Fig.~\ref{fig:res:1} (\textit{Top}), but the injected relation contains a stochastic component, as given by \eqref{eq:y:4}. Note that $f_{\rm true}$ is still standing for the deterministic part of the relation, as given by \eqref{eq:y:1}.}
    \label{fig:res:2}
\end{figure}

\begin{figure}[t]
    \centering
    \includegraphics[width=\linewidth]{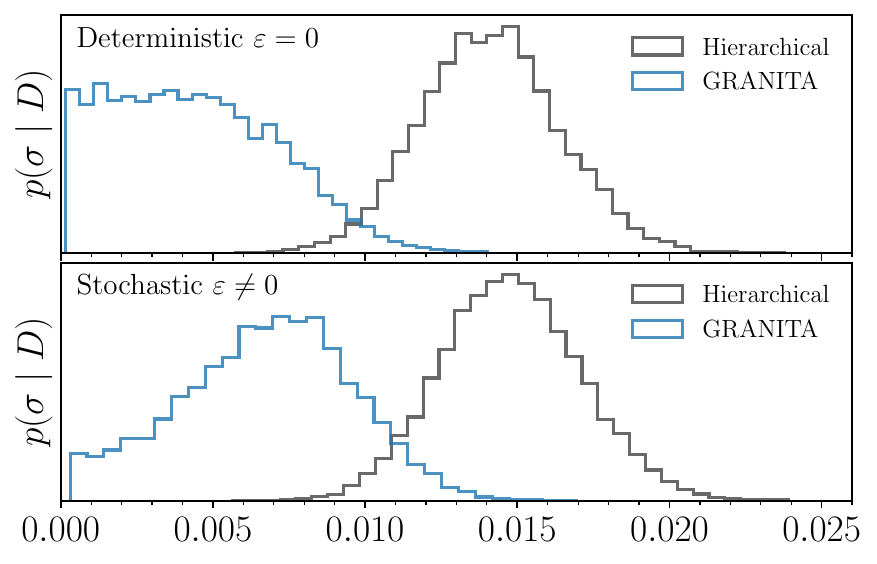}
    \caption{Reconstructed standard deviations for the \granita and for the hierarchical model, in the case of a deterministic underlying dependence (\textit{Top}) and with the addition of a stochastic component (\textit{Bottom}).}
    \label{fig:res:3}
\end{figure}

Figure \ref{fig:res:2} is the analogue of Fig.~\ref{fig:res:1} (\textit{Top}), in the case where $\delta y$ is generated according to Eq.~\eqref{eq:y:4}. Again, we see that the posterior sample functions are consistent with the injected $f_{\rm true}$ within the 95\% confidence band. 
Further insights is obtained by inspecting the posterior distribution of the free parameter $\sigma$. 
As shown in Fig.~\ref{fig:res:3}, $p(\sigma)$ largely supports $\sigma=0$ when the underlying functional dependence is deterministic as given by Eq.~\eqref{eq:y:1}, while $\sigma$ is constrained away from $0$ in the case of Eq.~\eqref{eq:y:4}. This illustrates how the parameter $\sigma$ in \granita is also endowed with physical meaning, since it informs us of whether the observed functional dependence is deterministic; 
in other words, $\sigma$ is suitable to capture unmodeled dependencies, such as the presence of environmental effects of astrophysical origin or the existence of primary BH hair.

By comparison, $\sigma$ (along with $\mu$) in the hierarchical method is a valuable summary statistic to detect the presence of deviations from GR. 
However, since its role is to account for all the variance about the global mean, it is incompatible with 0 in both cases.

\noindent{\bf \em Application to real data~--~}
We consider GW data from GR tests performed within the \pseob testing framework \cite{Ghosh:2021mrv} -- a test aiming at performing BH spectroscopy by incorporating pre-merger information. We consider a set of 10 GW events that meet the selection criteria detailed in \cite{LIGOScientific:2021sio}.

\begin{figure}[!th]
    \centering
    \includegraphics[width=\linewidth]{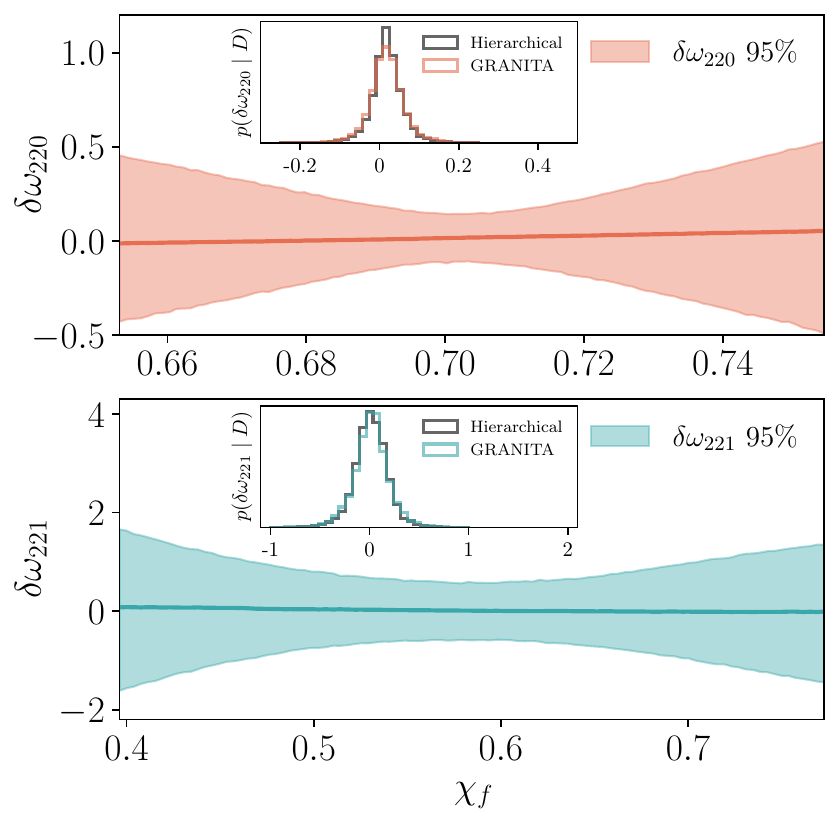}
    \caption{ \textit{Top} (\textit{Bottom}): the reconstructed functional dependence of $\delta \omega_{220}$ ($\delta \omega_{221}$) on $\chi_f$, obtained using the GWTC-3 posteriors from the \pseob (\pyring) analysis. Solid coral (teal) curve and shaded area denote the posterior median and $95\%$ intervals over the predictive range for $\chi_f$, respectively. \textit{Inset}: a comparison between posterior predictives obtained with \granita and the hierarchical analysis.}
    \label{fig:pseob:omega}
\end{figure}

We apply \granita to characterize a dependence between the deviation parameter $\delta\omega_{220}$ and the final BH spin $\chi_f$ -- i.e., we set $\theta_{\rm GR}=\chi_f$ and $\delta y=\delta\omega_{220}$. We choose $\delta\omega_{220}$ as it is the best constrained deviation parameter, compared to the large uncertainties affecting the recovery of $\delta\tau_{\rm 220}$ (see, e.g., Fig.~14 in \cite{LIGOScientific:2020tif}).

Note that, while in general deviation parameters in the BH ringdown depend on both the final mass $M_f$ and final spin $\chi_f$, the dependence on $M_f$ is suppressed for certain scalar-tensor theories, e.g., in the presence of Lorentz violation \cite{Barausse:2013nwa}. Moreover, \cite{Maselli:2023khq} shows that these theories exhibit ringdown deviations with the most promising prospects for being constrained by next-generation GW detectors, such as the Einstein Telescope \cite{Punturo:2010zz} and the Cosmic Explorer \cite{dwyer2015gravitational}. Therefore, our choice of parameters is dictated by a combination of both simplicity and physical motivation.

Despite the small number of events, the individual recoveries of the final BH spins are concentrated in a region between $\chi_f\approx0.5$ and $\chi_f\approx0.9$. Therefore, we choose unequally spaced nodes with locations $\boldsymbol{X}_{\rm nodes}=\{0,0.5,0.65,0.8,1.0\}$ and fix $l=0.5$. The extrema at $\chi_f=0$ and $\chi_f=1$ allow us to quantify the prior predictive uncertainty also outside of the data support for $\chi_f$.

For each event, the deviation parameters are sampled uniformly within ranges specificied in
the supplemental material.
On the other hand, since the analysis considers flat priors on the binary component masses and isotropic uniform priors on the component spins, a non-flat prior on $\chi_f$ is induced. We interpolate the induced prior $\pi(\chi_f)$ using Gaussian KDE and take it into account when applying Eq.~\eqref{eq:likelihood:1}

Figure \ref{fig:pseob:omega} (\textit{Top}) displays the results, zooming in on the portion of the parameter space that is effectively constrained by the data. 
We do not display the individual-event contours, as they are only loosely informative because of their individual low SNR. The results are consistent with no deviation from GR, which is further confirmed by $\sigma$ being compatible with $0$ and bounded to $\sigma<0.05$ at 90~\% credibility.

We also apply \granita to the 21 events selected by \cite{LIGOScientific:2021sio} to perform GR tests from the \pyring analysis, where $\delta\omega_{221}$ is used as a deviation parameter. Due to higher uncertainties in the posterior contours, we find larger bounds in the 95\% confidence region as shown in Fig.~\ref{fig:pseob:omega} (\textit{Bottom}), as well as on $\sigma$ ($\sigma<0.18$ at 90\% credibility), but the results are again consistent with no deviation from GR.

More detailed plots for the posterior recoveries are available in 
the supplemental material
 for both the \pseob and the \pyring analyses.

\noindent{\bf \em Discussion~--~}
In this work, we presented \granita, an extension of current GR tests incorporating the functional relation between the deviation parameters and the ordinary source parameters. Our approach is non-perturbative, and it leverages the expressive power of Gaussian processes to model functional dependence in a theory-agnostic way. Although we presented a proof-of-concept application to real data using public LIGO-Virgo-KAGRA data from binary BH ringdowns, the method can be easily adapted to characterize deviations from GR in more general testing frameworks.
We expect it to be relevant for testing GR given the increased brightness and number of detections with upcoming LIGO-Virgo-KAGRA observations \cite{2018LRR....21....3A}, and next-generation GW detectors \cite{Gupta:2023lga}, such as the Einstein Telescope \cite{Punturo:2010zz,Abac:2025saz}, Cosmic Explorer \cite{dwyer2015gravitational,Evans:2021gyd}, and LISA \cite{LISA:2024hlh}.

The approach leverages the flexibility of fixing the number and locations of the nodes, as well as the correlation length, allowing it to better adapt to the distribution of the observations within the parameter space. Moreover, the nodes and the correlation length can be optimized through Bayesian model selection, a refinement that we plan to introduce in future applications.

In our first applications, we restricted ourselves to a two-dimensional parameter space. More intricate correlations can be explored considering multidimensional spaces, as demonstrated by \cite{Zhong:2024pwb,2022PhRvD.105h4062S,2024PhRvD.109d4036D}.\footnote{Incidentally, we note that \cite{Zhong:2024pwb} also considers the relation between deviation parameter and GR parameters, in the context of inspiral-merger-ringdown consistency tests. Their analysis is fundamentally different from ours, in that they check for the presence of linear correlations, while we seek to reconstruct the proper functional relation.}Moreover, previous analyses have shown that biases in GR tests could arise when the entire population of source parameters is not modeled \cite{Payne:2023kwj} and when selection effects are not adequately included \cite{Essick:2023upv}. Selection effects are predominantly determined by the masses of the binary, especially for low SNR signals \cite{1993PhRvD..47.2198F}; therefore their exclusion is not likely to appreciably bias our analysis, as we restricted to dependencies on the final spin. However, as more GW events are detected and constraints on the deviation parameters become increasingly precise, it will be crucial to consider the full parameter space together with selection effects.

Finally, our approach naturally complements ongoing efforts to model gravitational-wave effects non-perturbatively in theories beyond GR \cite{Julie:2024fwy, Chung:2024ira, Cano:2024ezp, Blazquez-Salcedo:2024oek}: nodes can be defined to span fluctuations about explicit beyond-GR solutions. This can be implemented by setting a nonvanishing expectation value for the Gaussian process (as shown in \cite{Pozzoli:2023lgz}) rather than starting from vanishing ones as in this work. Such a scheme would allow us to explore deviations from a baseline theory, thus effectively augmenting the space of beyond-GR models.

{\em Acknowledgements~--~}
We are grateful to D. Gerosa, A. Renzini, and C.~J. Moore for useful comments and valuable inputs.
We thank A.~Maselli for clarifications and discussions.
C.P. is supported by 
ERC Starting Grant No.~945155--GWmining, 
Cariplo Foundation Grant No.~2021-0555, 
MUR PRIN Grant No.~2022-Z9X4XS, 
MUR Grant ``Progetto Dipartimenti di Eccellenza 2023-2027'' (BiCoQ),
and the ICSC National Research Centre funded by NextGenerationEU. 
C.P. and R.B. acknowledge support from the ICSC National Research Center funded by NextGenerationEU.
R.B. acknowledges support by the Italian Space Agency grant Phase B2/C activity for LISA mission, Agreement n.2024-NAZ-0102/PER.
This material is based upon work supported by NSF's LIGO Laboratory which is a major facility fully funded by the National Science Foundation, and has made use of data or software obtained from the Gravitational Wave Open Science Center (gwosc.org), a service of the LIGO Scientific Collaboration, the Virgo Collaboration, and KAGRA.

\textit{Software}:
We acknowledge usage of the following 
\textsc{Python}~\cite{10.5555/1593511} 
packages for modeling, analysis, post-processing, and production of results throughout: \textsc{corner}~\cite{corner}, \textsc{jax}~\cite{frostig2018compiling,bradbury2021jax}, \textsc{matplotlib}~\cite{2007CSE.....9...90H}, \textsc{numpyro}~\cite{phan2019composable,bingham2019pyro}.
The data supporting this work are made available upon reasonable request by the authors.

\bibliographystyle{apsrev4-2} %
\bibliography{main}%

\appendix
\input{appendixcontent}

\end{document}

%% file: appendixcontent.tex
\clearpage
\onecolumngrid

\section{SUPPLEMENTAL MATERIAL: Analyses details}
\label{app:a}
We briefly summarize here some technical details on how we performed the analysis on the toy and GWTC-3 data shown in
the main text.
To sample the likelihood in~\cref{eq:likelihood:1} we employ No-U-Turn sampling (NUTS)~\cite{2011arXiv1111.4246H}, a variant of Hamiltonian Monte Carlo (HMC)~\cite{2011hmcm.book..113N}, as implemented in the \textsc{numpyro} package~\citep{phan2019composable,bingham2019pyro}. In doing so, we leverage likelihood gradients defined through automatic differentiation.
We instantiate each model and sample parameter posteriors with $5000$ warmup steps and $10000$ sampling steps, validating each chain convergence through Gelman-Rubin statistics~\cite{1992StaSc...7..457G} and thinning it to suitably uncorrelated posterior samples.

\subsection{Toy model}
We choose the following priors for the parameterized toy-model defined in~\cref{eq:y:3}: $A,B$ are sampled from a standard normal, while $C$ is sampled from a half-normal distribution with scale parameter equal to $1$; $\mu_X$, $\sigma_X$, and $\sigma$ are sampled uniformly in $[0,1]$. 
In the \granita toy-model analysis we assume the same priors over the shared parameters, while we assume the nodes $\boldsymbol{Y}$ to be i.i.d. following a standard normal.

\subsection{GWTC-3 events}
\label{app:gwtc3} 

\begin{figure*}[!th]
\centering
\includegraphics[width=0.9\textwidth]{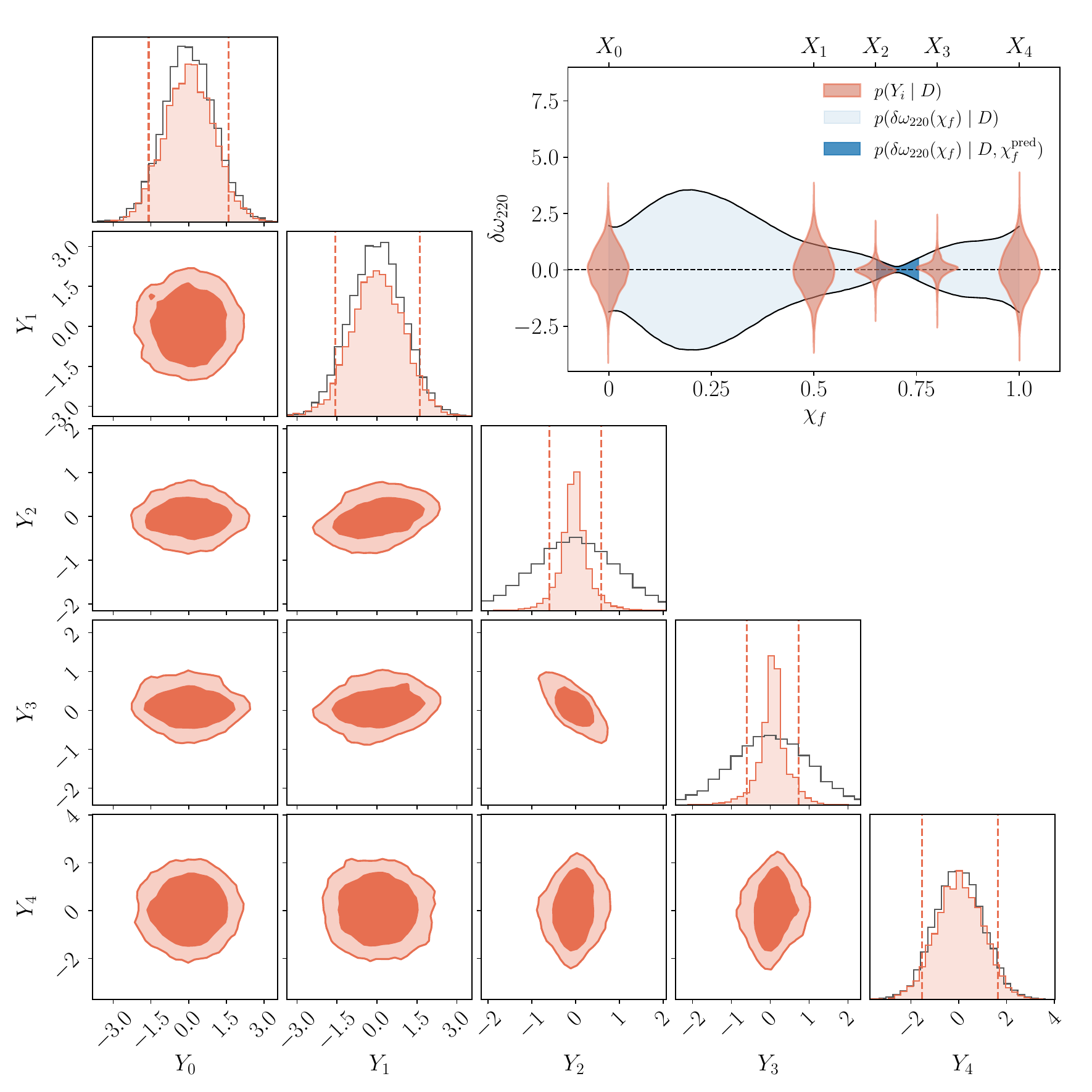}
\caption{Posterior distribution for $\delta\omega_{220}$ from the selection of \pseob-analyzed events in the GWTC-3 testing GR paper \cite{LIGOScientific:2021sio}. (\textit{Main panel}) corner plot showing the posterior on nodes $Y_i$, as introduced in the main text, and marginalized over $\sigma_X$, $\mu_X$, $\sigma$. Shaded regions in 2D densities denote $68\%$ and $90\%$ confidence regions.
Vertical dashed lines in 1D histograms denote equal-tailed $90\%$ confidence intervals. For reference, grey histograms denote the prior assumed for each node. (\textit{Inset}) Dark orange violins denote the marginal posteriors on $Y_i$, positioned at their chosen node-position $X_i$. The light blue shaded area denotes the  posterior $95\%$ confidence region on function $\delta\omega_{220}(\chi_f$), as inferred with \granita over the interval $[0,1]$ for $\chi_f$. In darker blue, the same quantity is highlighted over the range yielding the $95\%$ posterior support for $\chi_f$, which we focus on in the top panel of~\cref{fig:pseob:omega}.}\label{fig:corner-plot-real-pseob}
\end{figure*}

\begin{figure*}[!th]
\centering
\includegraphics[width=0.9\textwidth]{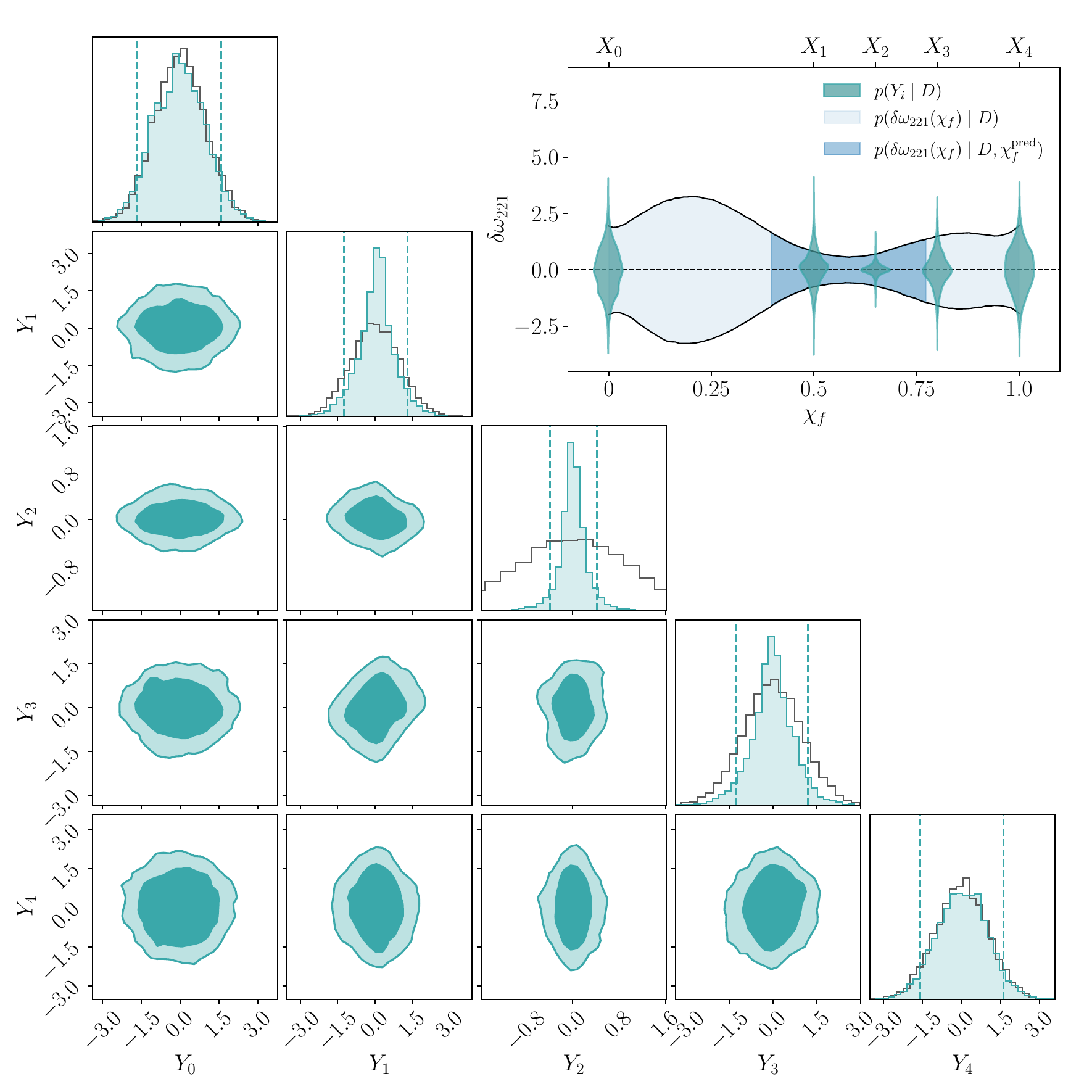}
\caption{Posterior distribution for $\delta\omega_{221}$ from the selection of \pyring-analyzed events in the GWTC-3 testing GR paper \cite{LIGOScientific:2021sio}. (\textit{Main panel}) corner plot showing the posterior on nodes $Y_i$, as introduced in the main text, and marginalized over $\sigma_X$, $\mu_X$, $\sigma$. Shaded regions in 2D densities denote $68\%$ and $90\%$ confidence regions.
Vertical dashed lines in 1D histograms denote equal-tailed $90\%$ confidence intervals. For reference, gray histograms denote the prior assumed for each node. (\textit{Inset}) Teal violins denote the marginal posteriors on $Y_i$, positioned at their chosen node-position $X_i$. The light blue shaded area denotes the  posterior $95\%$ confidence region on function $\delta\omega_{221}(\chi_f$), as inferred with \granita over the interval $[0,1]$ for $\chi_f$. In darker blue, the same quantity is highlighted over the range yielding the $95\%$ posterior support for $\chi_f$, which we focus on in the bottom panel of~\cref{fig:pseob:omega}.}\label{fig:corner-plot-real-pyring}
\end{figure*}

For the inferences on events from GWTC-3, nodes $\boldsymbol{Y}$ are \emph{a priori} i.i.d. following a standard normal distribution; $\sigma$ is uniformly distributed in $[0,0.53]$ and $[0,1.54]$ for the \pseob and \pyring analysed events, respectively; upper-bounds were chosen for computational convenience, however posteriors are likelihood-dominated and therefore largely unaffected by our choice. Parameters describing the model stochastic component, i.e. $\mu_X, \sigma_X$, are uniformly distributed in $[0,1]$ for the \pseob and \pyring analysis.
In~\cref{fig:corner-plot-real-pseob,fig:corner-plot-real-pyring} we show the joint posteriors on $\boldsymbol{Y}$ to illustrate correlations.
First, we see that in regions with data support, the posterior recovery of $\boldsymbol{Y}$ is significantly more constrained than the prior; at the same time, we essentially recover the prior in regions with negligible observational coverage, as expected. The violin plots in the insets display the posterior of the nodes in correspondence with their coordinates $\boldsymbol{X}$. We highlight in dark blue the 95\% posterior support for $\chi_f$, which is representative of the observationally most covered region.

